\documentclass[10pt,conference]{IEEEtran}

\usepackage{cite}

\ifCLASSINFOpdf
   \usepackage[pdftex]{graphicx}
   \graphicspath{{figs/}}
   \DeclareGraphicsExtensions{.pdf,.jpeg,.png}
\else
   \usepackage[dvips]{graphicx}
   \graphicspath{{../figs/}}
   \DeclareGraphicsExtensions{.eps}
\fi
\usepackage[cmex10]{amsmath}
\usepackage{amsthm}

\usepackage{algorithmic}

\usepackage{array}

\ifCLASSOPTIONcompsoc
  \usepackage[caption=false,font=normalsize,labelfont=sf,textfont=sf]{subfig}
\else
  \usepackage[caption=false,font=footnotesize]{subfig}
\fi
\usepackage{url}

\usepackage{comment}
\usepackage[hidelinks]{hyperref}
\usepackage{booktabs}
\usepackage{xcolor}

\newif\iffinal
\finaltrue
\newcommand{\cmtid}{99999}

\iffinal
\else
\usepackage[switch]{lineno}
\fi

\begin{document}
%
\title{Exploring Foundation Models for Synthetic Medical Imaging: A Study on Chest X-Rays and Fine-Tuning Techniques}



\iffinal

\author{\IEEEauthorblockN{Davide Clode da Silva, Marina Musse Bernardes, Nathália Giacomini Ceretta, Gabriel Vaz de Souza,\\Gabriel Fonseca Silva, Rafael Heitor Bordini and Soraia Raupp Musse}
\IEEEauthorblockA{Pontifical Catholic University of Rio Grande do Sul (PUCRS), RS, Brazil\\
Email: gabriel.fonseca94@edu.pucrs.br, soraia.musse@pucrs.br}}

\else
  \author{SIBGRAPI Paper ID: \cmtid \\ }
  \linenumbers
\fi

\maketitle

\begin{abstract}
Machine learning has significantly advanced healthcare by aiding in disease prevention and treatment identification. However, accessing patient data can be challenging due to privacy concerns and strict regulations. Generating synthetic, realistic data offers a potential solution for overcoming these limitations, and recent studies suggest that fine-tuning foundation models can produce such data effectively. 
In this study, we explore the potential of foundation models for generating realistic medical images, particularly chest x-rays, and assess how their performance improves with fine-tuning. We propose using a Latent Diffusion Model, starting with a pre-trained foundation model and refining it through various configurations. Additionally, we performed experiments with input from a medical professional to assess the realism of the images produced by each trained model.
\end{abstract}


\IEEEpeerreviewmaketitle

\newcommand\red[1]{{#1}}

\section{Introduction}

In recent years, Machine Learning (ML) has played a crucial role in healthcare. For instance, in disease prevention and treatment, ML can help analyze large datasets to identify trends and predict disease outcomes, such as modeling the progression and treatment of cancerous conditions \cite{kourou2015machine}. 
Nevertheless, the adoption of ML techniques in healthcare has been slow due to factors like scarcity of patient data, data privacy concerns, regulatory requirements, and the critical nature of healthcare decisions \cite{goncalves2020generation, thapa2021precision}.


One of the main barriers to accessing patient medical data is the need to protect sensitive and confidential information from unauthorized access. Additionally, the lack of standardized records and the effort required to collect medical data pose significant challenges \cite{KESHTA2021177, almaghrabi2022patient}. 
Generating synthetic, realistic, and high-quality medical data could be a viable alternative to mitigate some of these issues. The industry predicts a significant increase in the availability of synthetic data in the coming years, potentially doubling the amount of real data currently available~\cite{sundaram2021ganbased}.
Examples include generative models for creating photorealistic images from natural language descriptions \cite{reed2016generative} and improving GAN-based (Generative Adversarial Network) models' performance on super-resolution images \cite{dou2020pca}.

In the context of medical images, generative models may offer a practical solution, with some work focusing on fine-tuning foundation models using small datasets~\cite{azad2023foundational}. Foundation Models (FMs) are machine learning models trained on a wide range of generalized and unlabeled data, typically using self-supervision techniques. Examples of Foundation Models include ELMo~\cite{peters2018deep}, 
GPT-3~\cite{brown2020language}, CLIP~\cite{radford2021learning}, ResNet~\cite{he2016deep}, DALL-E~\cite{ramesh2021zero}, and Stable Diffusion~\cite{rombach2022high}. These models have achieved significant advancements in various complex tasks~\cite{bommasani2021opportunities}, such as Question Answering~\cite{brown2020language}, Knowledge Base Construction~\cite{petroni2019language}, and Information Retrieval~\cite{guu2020retrieval}. Fine-tuning such models targets the foundation model generalization capabilities into specific applications.

One example is the MedCLIP project \cite{wang2022medclip}, where a Contrastive Language-Image Pre-Training (CLIP) model was adapted using the ROCO dataset~\cite{radford2021learning}, which includes fine-tuned images and corresponding captions. Researchers claim that the adjusted model, MedCLIP, can identify higher-level characteristics such as the image modality, distinguishing between PET (Positron Emission Tomography) scans and ultrasound scans. 
Other studies have used generative model approaches for various purposes, including synthesizing realistic medical data \cite{motamed2021data, pinaya2022brain} and augmenting datasets for deep learning model training \cite{dou2020pca}.

In this work, we present an initial exploration of the capabilities of foundational models on generating realistic medical images and how their performance is impacted by fine-tuning. \red{We use a small dataset during the fine-tuning process to assess the ability of FMs to learn effectively with limited data}. 
We focus on generating chest x-ray images, considering healthy and unhealthy diagnoses. We propose utilizing a Latent Diffusion Model (LDM) approach, employing a pre-trained foundation model as a basis and later fine-tuning it with different configurations.  Furthermore, we conducted experiments with the assistance of a medical professional to evaluate the realism of images generated by each model trained.


The remainder of this work is organized as follows: Sections~\ref{sec:related_work} presents related work that has played an important role in understanding different approaches to generating synthetic medical data. Section~\ref{sec:proposed_model} details our proposed method for generating synthetic medical images. Section~\ref{sec:prelimirany_results} presents the experiments conducted and preliminary results achieved. Finally, Section~\ref{sec:final_considerations} presents our final considerations, limitations, and possible future contributions.

\section{Related Work}
\label{sec:related_work}

The literature presents various image-generated techniques, often based on a textual description. 
Chen et al~\cite{chen2022re} proposed Re-Imagen (Retrieval-Augmented Text-to-Image Generator), a generative model that utilizes retrieved information to produce accurate images. It is distinguished by its capability to generate faithful images, even for rare or never-before-seen entities. 
Zhou et al.~\cite{zhou2022towards} introduced LAFITE (LAnguage-Free traIning for Text-to-image gEneration) to address one of the major challenges in training text-to-image generation models: the requirement for a large number of image-text pairs. The authors claim that their proposed model can be trained without any text data, and it has shown promising results.

Focusing on medical images, Pinaya et al. \cite{pinaya2022brain} proposed a model to generate synthetic images from high-resolution brain MRIs. Their model uses data to learn the probabilistic distribution of brain images based on covariates like age, sex, and brain structure volumes. The authors employed Latent Diffusion Models (LDM), which combine autoencoders to compress input data into a lower-dimensional latent representation with the generative modeling properties of diffusion models. The compression model, crucial for enabling the scalability of high-resolution medical images, was trained using a combination of perceptual loss and an adversarial objective based on patches—specific small areas of an image used to modify or manipulate parts of the image to create adversaries or deceive machine learning algorithms.

Ali et al. \cite{ali2022spot} investigated medical image synthesis using diffusion models. Initially, the authors used a pre-trained DALLE2~\cite{ramesh2021zero} model to generate lung x-ray and CT images from text prompts, then, they trained a stable diffusion model on 3,165 x-ray images. For evaluation, two independent radiologists conducted a qualitative analysis by labeling randomly chosen samples as real, fake, or uncertain. The results indicated that the diffusion model could effectively translate features specific to certain medical conditions into chest X-rays or CT images.

Packhauser et al. \cite{packhauser2023generation,Fav2017} utilized (LDM) to generate high-quality chest x-ray images while preserving the privacy of sensitive biometric information. 
Conditional information was embedded using a trainable lookup table combined with cross-attention at the U-Net bottleneck. The model was trained on a dataset of chest X-rays from 30,805 patients, including metadata with 14 abnormality labels and an additional class for healthy individuals. The generated dataset was evaluated in a thoracic abnormality classification task, and the approach outperformed GAN-based methods.

Other recent work focuses on image generation for dataset augmentation. 
Sundaram et al. \cite{sundaram2021ganbased} proposed using Generative Adversarial Networks (GANs) to augment a chest x-ray dataset to address class imbalance. Their strategy involved creating synthetic chest X-ray images featuring at least one of three underrepresented pathologies: lung injury, pleural injury, or fracture. These synthetic images were added to the original dataset to reduce class imbalance. 
Motamed et al. \cite{motamed2021data} introduced the Inception Augmentation GAN (IAGAN) for chest x-ray data augmentation, targeting semi-supervised detection of pneumonia and COVID-19. Inspired by the Data Augmentation Generative Adversarial Networks (DAGAN) \cite{antoniou2017data}, their model aimed to generate synthetic data to enhance training datasets for other models.

\section{Proposed Model}
\label{sec:proposed_model}

In this work, our main task is to perform the fine-tuning of a Latent Diffusion Model in order to generate high-resolution synthetic chest x-ray images. Section~\ref{sec:dataset} describe the dataset utilized for fine-tuning, composed of real chest x-ray images. Section~\ref{sec:fine_tuning} describes the fine-tuning process, resources and parameterization utilized.

\subsection{Dataset}
\label{sec:dataset}

In this study, we use the Montgomery County CXR Set dataset for tuberculosis, developed by the National Library of Medicine in collaboration with the Department of Health and Human Services in Montgomery\footnote{Information on the research is available at: \url{https://lhncbc.nlm.nih.gov/LHC-downloads/dataset.html}}. 
The dataset is publicly available\footnote{The dataset is available for download at: \url{https://data.lhncbc.nlm.nih.gov/public/Tuberculosis-Chest-X-ray-Datasets/Shenzhen-Hospital-CXR-Set/index.html}} and is composed of 138 postero-anterior chest x-ray images. Of these, 80 are from normal (or healthy) cases, and 58 are from abnormal (or unhealthy) cases with manifestations consistent with tuberculosis. 

\begin{figure}[!ht]
  \centering
  \subfloat[``Normal chest x-ray'']{\includegraphics[width=0.30\textwidth]{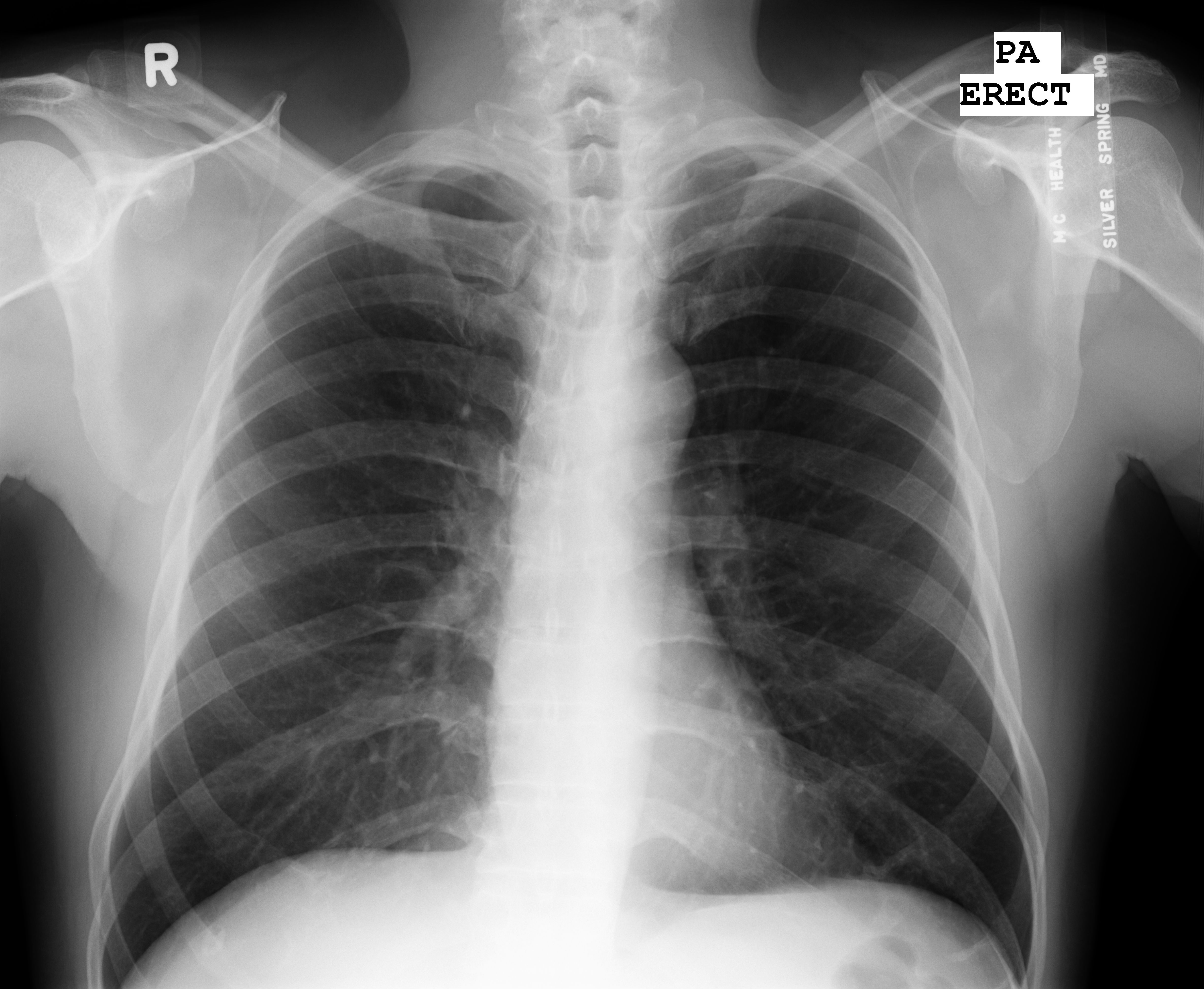}\label{fig:sample_normal}}
  \hfill
  \subfloat[``Large infiltrate Right Upper Lobe with cavitation plus infiltrate in RML. Consistent with active cavitary TB.'']{\includegraphics[width=0.30\textwidth]{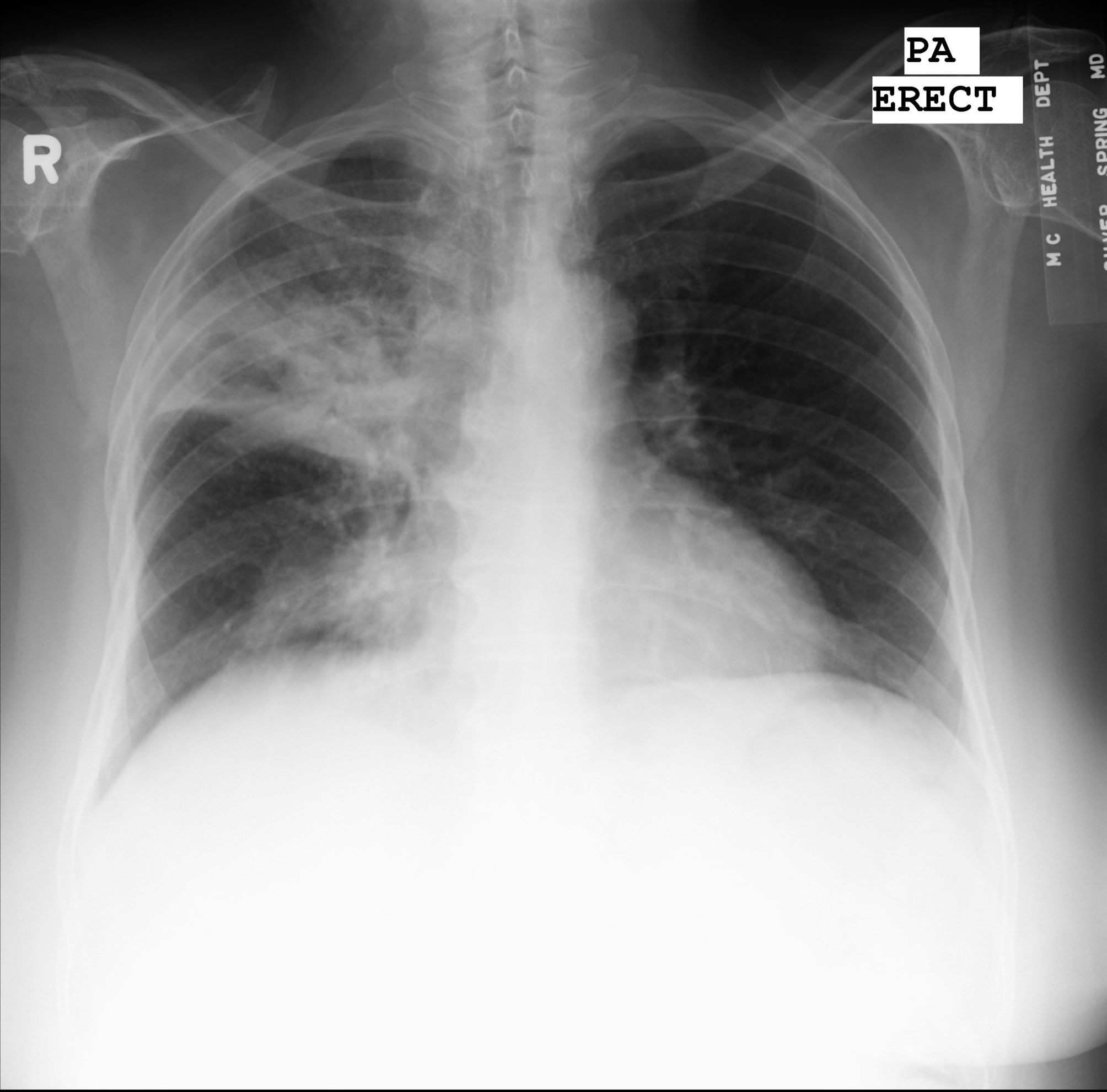}\label{fig:sample_abnormal}}
  \caption{Sample images from the dataset and their respective annotations~\cite{jaeger2014two}. (a) presents a normal case, and (b) an abnormal case.}
  \label{fig:dataset_sample}
\end{figure}

Additionally, the dataset includes consensus annotations from two radiologists for resized 1024 × 1024 images and a report describing the imaging results~\cite{jaeger2014two}. Figure~\ref{fig:dataset_sample} presents sample images of both cases and their respective annotations. All images are anonymized and available in PNG format.
Based on the 138 x-rays available, we used a subset consisting of 30 images (50\% healthy images and 50\% unhealthy images) to fine-tune the models. We chose to work with a data set of this size as an initial exploration so that we could obtain more precise guidance on the next steps to be taken in future work. 

\subsection{Fine-tuning}
\label{sec:fine_tuning}

In our approach, we utilized the Kohya-ss GUI\footnote{Available at: \url{https://github.com/bmaltais/kohya_ss}}, a friendly user interface that allows the setup, training and fine-tuning of diffusion models. 
The interface provides different techniques for fine-tuning, such as DreamBooth ~\cite{ruiz2023dreambooth} and LoRA (Low-Rank Adaptation)~\cite{hu2021lora}. 
We believe LoRA to be a more appropriate option for our method, given its training efficiency by having a smaller number of parameters, reduced hardware requirements for adaptive optimizers, versatility in fine-tuning, and a smaller final model~\cite{hu2021lora}. Additionally, we utilized TensorBoard\footnote{Available at: \url{https://github.com/tensorflow/tensorboard}} alongside the Kohya-ss GUI as a visualization toolkit for the loss and learning rates during the fine-tuning.

We selected \textit{``stable-diffusion-v1-5''}
, proposed by Rombach et al.~\cite{rombach2022high}, as our foundation model, which is capable of generating realistic images based on textual descriptions. 
We performed a fine-tuning on this foundation model utilizing the chest x-ray dataset described in Section~\ref{sec:dataset}. For the fine-tuning process, we consider different optimizers, described next:
\begin{itemize}
    \item AdamW8bit: is a variant of Adam, a stochastic optimization method designed for large-scale machine learning problems. It adjusts the learning rate for each model weight individually and computes adaptive learning rates for different parameters~\cite{dettmers20218, 8_bit_optimizers, kingma2017adam}.
    \item Adafactor: aims to overcome the memory requirements of stochastic optimization methods like RMSProp and Adam by storing only the row and column sums of these averages, reducing memory usage while retaining adaptability~\cite{shazeer2018adafactor}.
    \item DAdaptSGD: is a variant of DAdaption, specifically designed to automatically determine the learning rate in the Stochastic Gradient Descent (SGD) optimization algorithm \cite{defazio2023learning}.
    \item Prodigy: is an optimizer that estimates the distance to the solution, allowing for optimal learning rate adjustment in adaptive methods like Adagrad and Adam. It's an adaptation of the D-Adaptation method for learning-rate-free learning, where the learning rate is set automatically \cite{mishchenko2023prodigy}.
\end{itemize}

We fine-tuned five distinct models, two utilizing the AdamW8bit, and one with each remaining optimizer. Table~\ref{tbl:configurations} presents the parameterization of each training. All models were trained for 100 epochs. 
Finally, we utilized the Stable Diffusion Web UI\footnote{Available at: \url{https://github.com/AUTOMATIC1111/stable-diffusion-webui}} for generating the images. The interface allows the user to select different models, image quantities and dimensions, seeds, and the prompt used for the generation.

\begin{table}[ht]
\centering
\caption{Parametrization of the fine-tuning process. All models were trained using \textit{``stable-diffusion-v1-5''} as foundation model. The Adam8bit optimizer was used in two distinct models, M1 using its default parameters, and M2 using a configuration similar to the remaining optimizers.
}
\resizebox{0.49\textwidth}{!}{
\begin{tabular}{|c|c|c|c|c|c|}
\hline
Model                    & M1          & M2          & M3          & M4          & M5          \\ \hline
Optimizer                & AdamW8bit   & AdamW8bit   & Adafactor   & DAdaptSGD   & Prodigy     \\ \hline
Clip Skip                & 2           & 2           & 2           & 2           & 2           \\ \hline
Epochs                   & 100         & 100         & 100         & 100         & 100         \\ \hline
LR                       & $1.10^{-4}$ & $1.10^{-4}$ & $1.10^{-4}$ & $1.10^{-4}$ & $1.10^{-4}$ \\ \hline
Max Resolution           & \red{512x512}     & \red{512x512}     & \red{512x512}     & \red{512x512}     & \red{512x512}     \\ \hline
LR Scheduler             & constant    & constant    & constant    & constant    & constant    \\ \hline
Train Batch Size         & 2           & 2           & 2           & 2           & 2           \\ \hline
Text Encoder LR          & $5.10^{-5}$ & $5.10^{-5}$ & $5.10^{-5}$ & $1.10^{-5}$ & $1.10^{-5}$ \\ \hline
Unet lr                  & $1.10^{-4}$ & $1.10^{-4}$ & $1.10^{-4}$ & $1.10^{-5}$ & $1.10^{-5}$ \\ \hline
VAE Batch Size           & 0           & 32          & 32          & 32          & 32          \\ \hline
Noise Offset Type        & Original    & Multires    & Multires    & Multires    & Multires    \\ \hline
Noise Discount  & 0           & 0.1         & 0.1         & 0.1         & 0.1         \\ \hline
Noise Iteration & 0           & 6           & 6           & 6           & 6           \\ \hline
\end{tabular}}
\label{tbl:configurations}
\end{table}

\section{Experimental Results}
\label{sec:prelimirany_results}

For our experiments, we generated images of chest x-rays using six different models: the pre-trained foundational model, named M0, and the five models fine-tuned using the different optimizers presented in Table \ref{tbl:configurations}, named M1 through M5. Each model was used to generate two sets of 12 images each: one set representing normal cases, created with the prompt \textit{``healthy or normal human chest x-ray''}; and the other representing abnormal cases, created with the prompt \textit{``Human chest x-ray with tuberculosis. Bilateral miliary nodules with Right Middle Lobe infiltrate. Right pleural effusion''}.

\begin{figure}[!ht]
  \centering
  \subfloat[M0-Normal (Foundation)]{\includegraphics[width=0.24\textwidth]{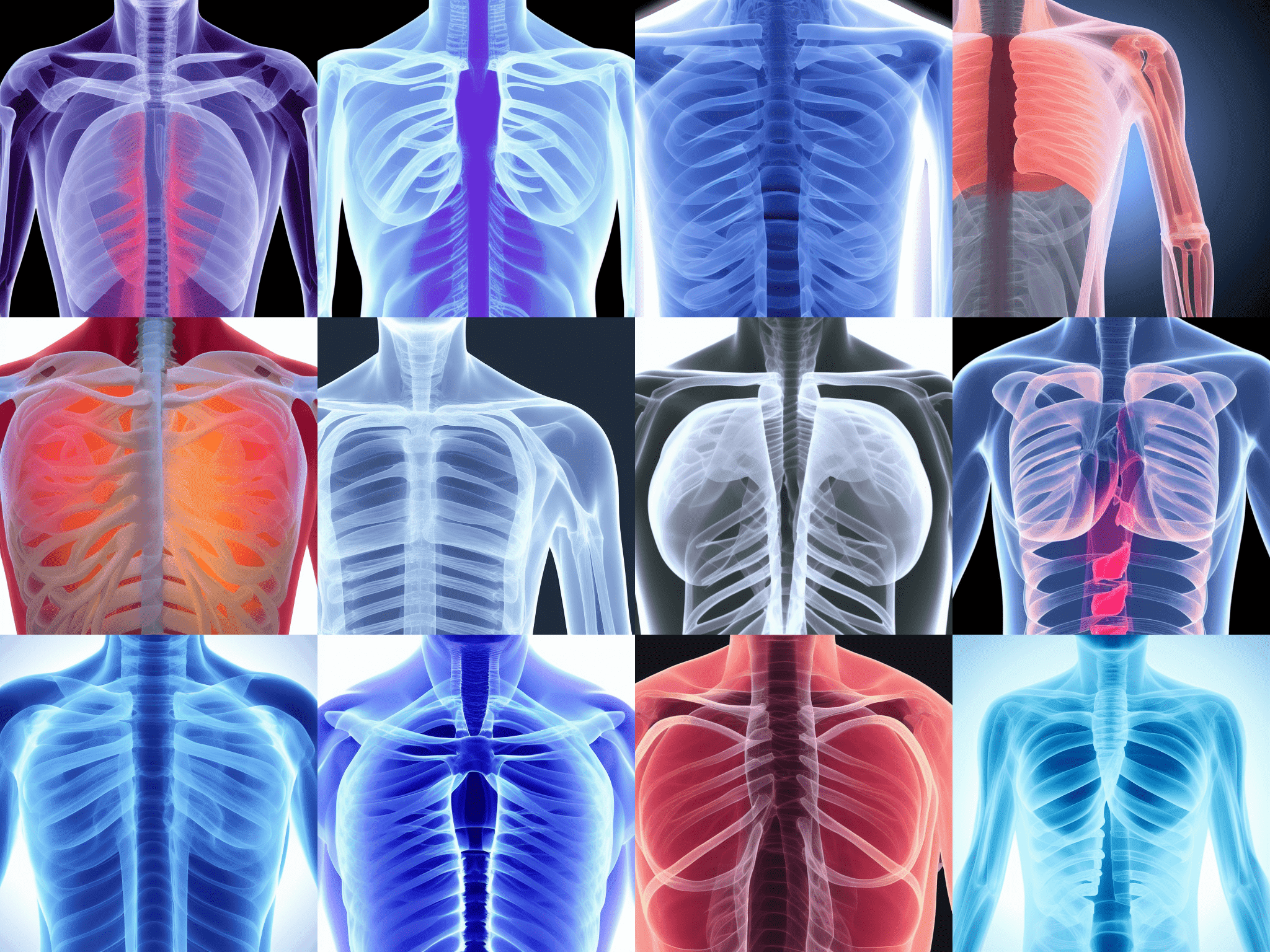}\label{fig:M0_normal}}
  \hfill
  \subfloat[M1-Normal (Adam8bit1)]{\includegraphics[width=0.24\textwidth]{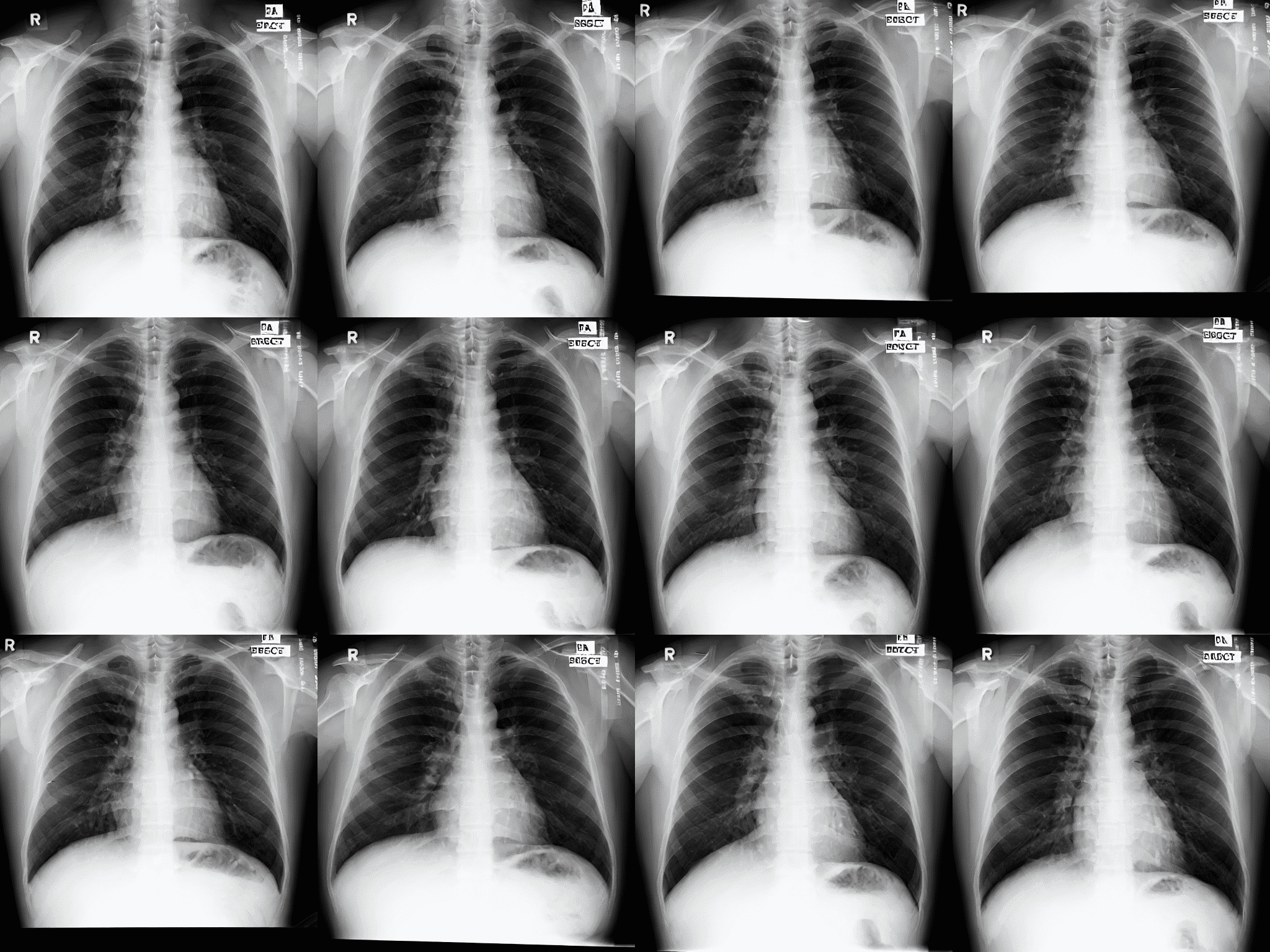}\label{fig:M1_normal}}
  \vfill
  \subfloat[M2-Normal (Adam8bit2)]{\includegraphics[width=0.24\textwidth]{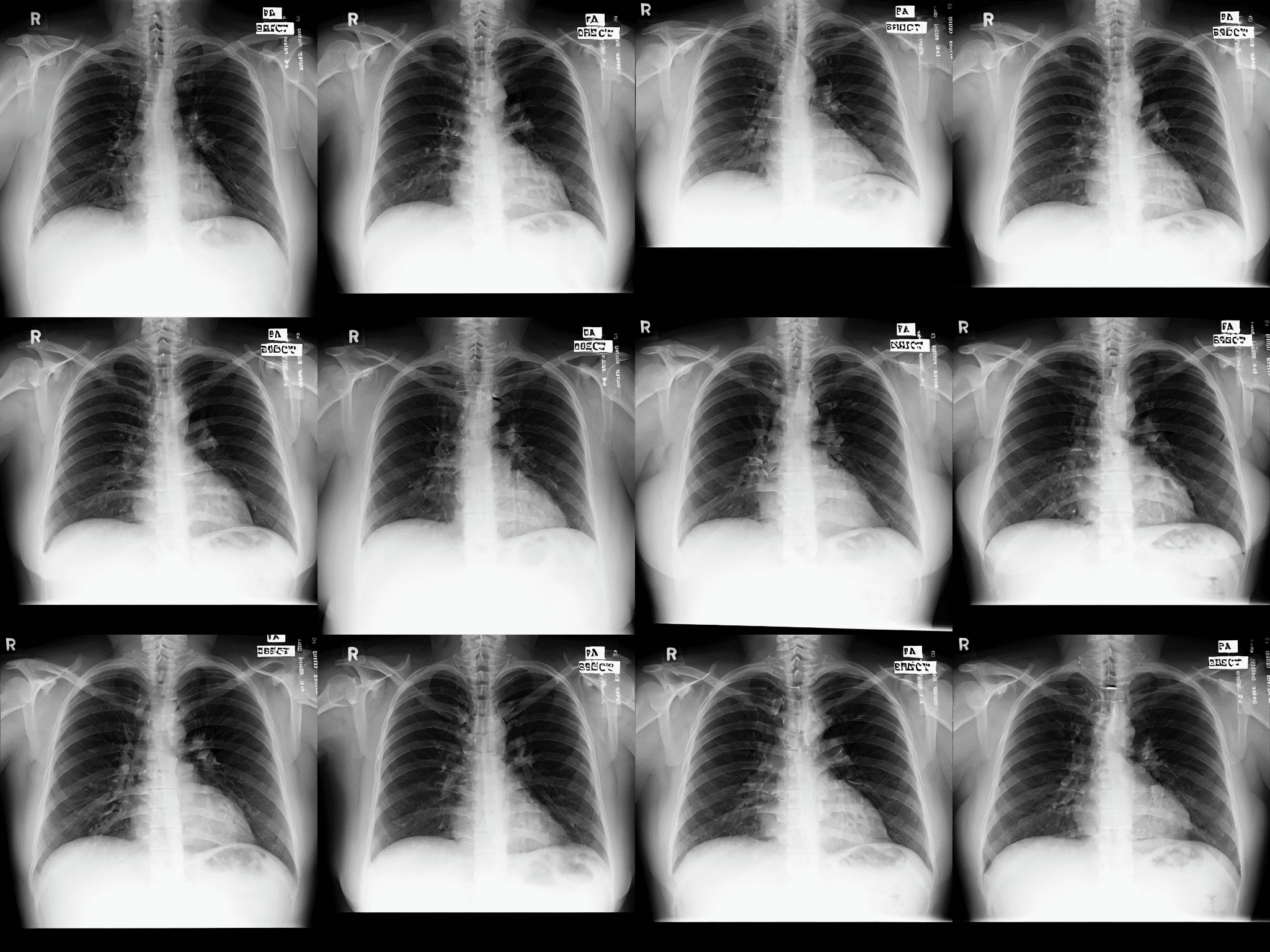}\label{fig:M2_normal}}
  \hfill
  \subfloat[M3-Normal (Adafactor)]{\includegraphics[width=0.24\textwidth]{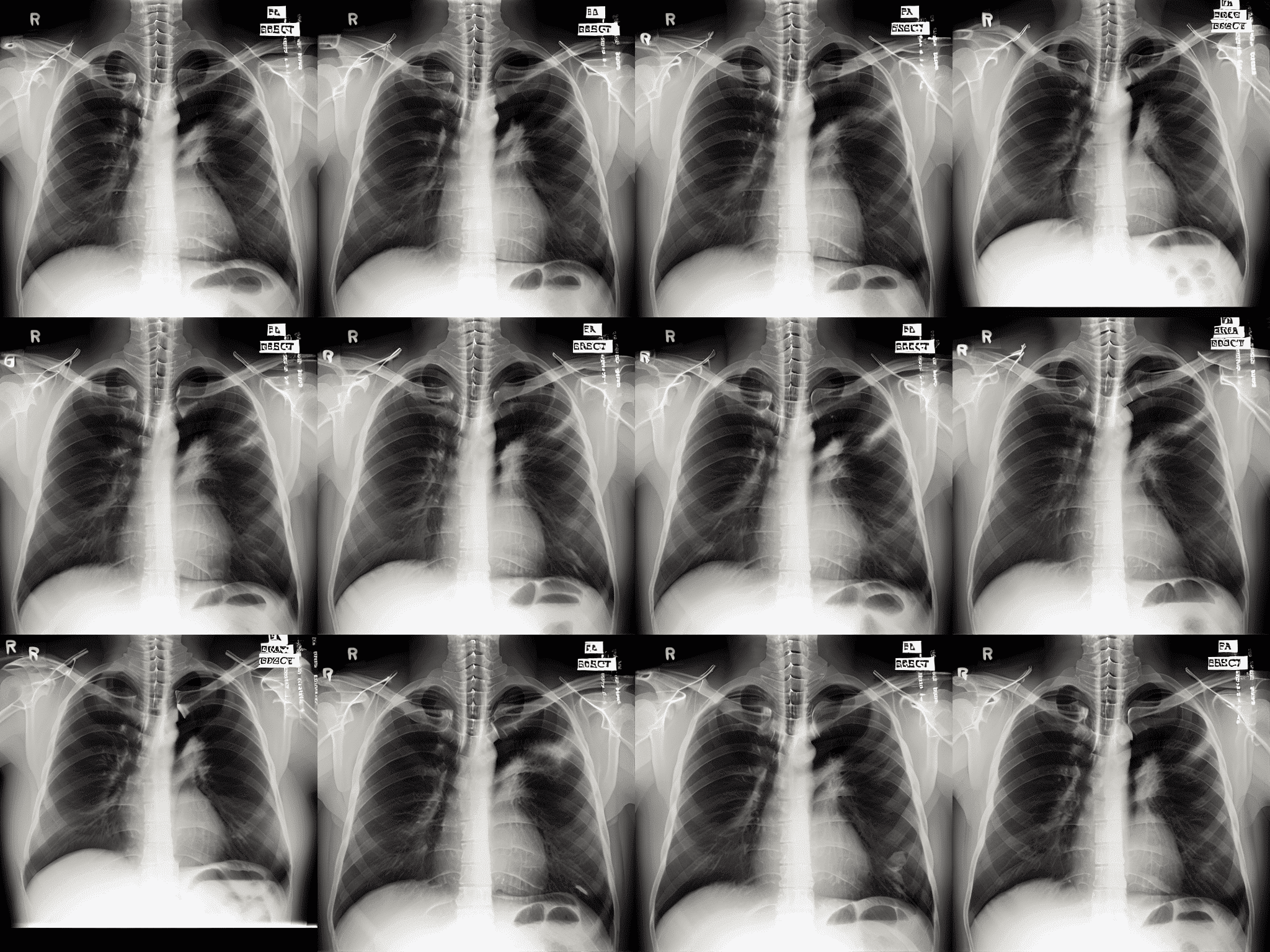}\label{fig:M3_normal}}
  \vfill
  \subfloat[M4-Normal (DAdaptSGD)]{\includegraphics[width=0.24\textwidth]{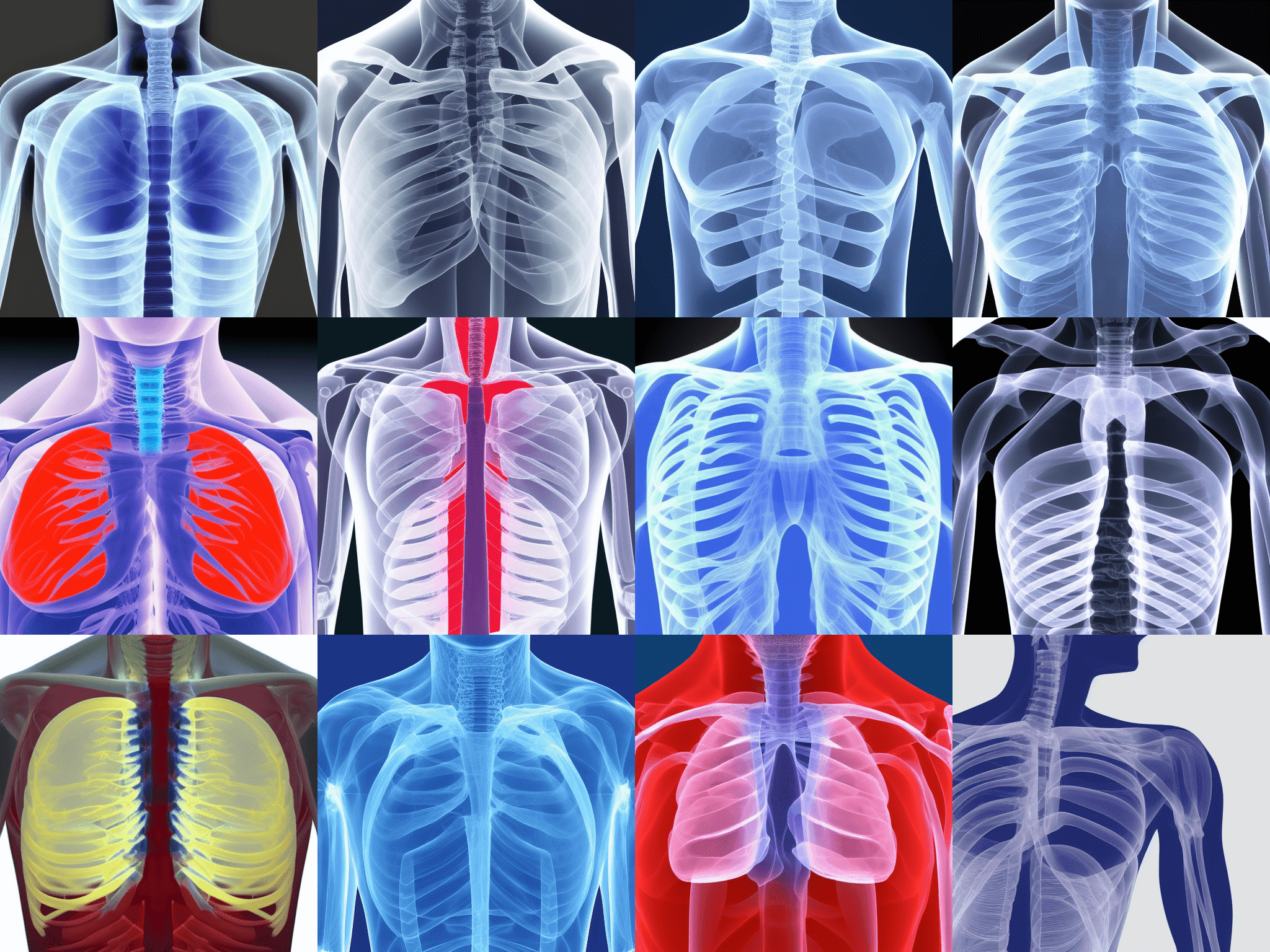}\label{fig:M4_normal}}
  \hfill
  \subfloat[M5-Normal (Prodigy)]{\includegraphics[width=0.24\textwidth]{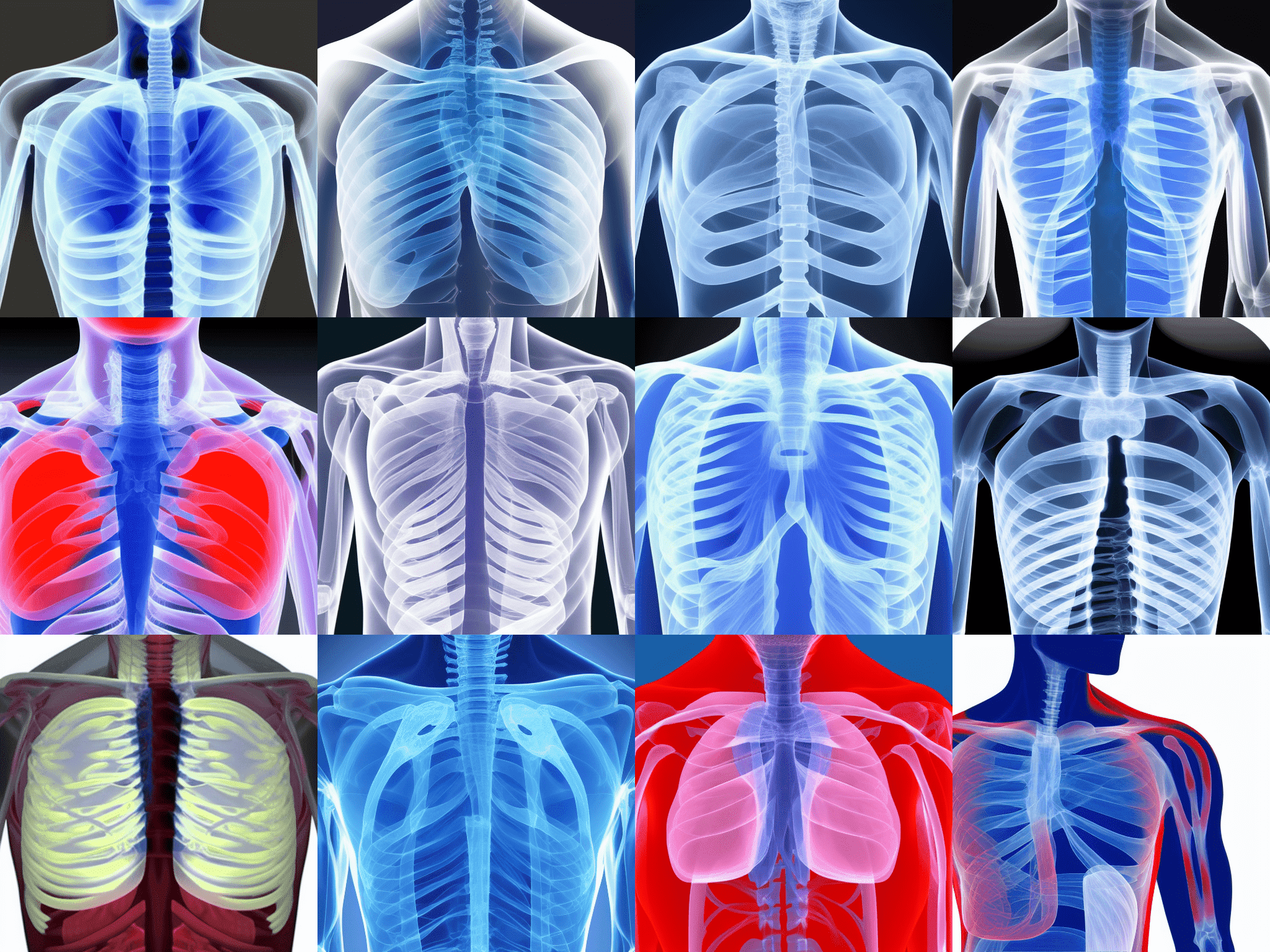}\label{fig:M5_normal}}
  \caption{Set of normal chest x-ray images generated by the models presented in Table \ref{tbl:configurations}. All models except M1 were fine-tuned for 100 epochs on a dataset of 30 chest x-ray images. All images were generated using the prompt \textit{"healthy or normal human chest x-ray"}.}
  \label{fig:normal_images}
\end{figure}

\begin{figure}[!ht]
  \centering
  \subfloat[M0-Abnormal (Foundation)]{\includegraphics[width=0.24\textwidth]{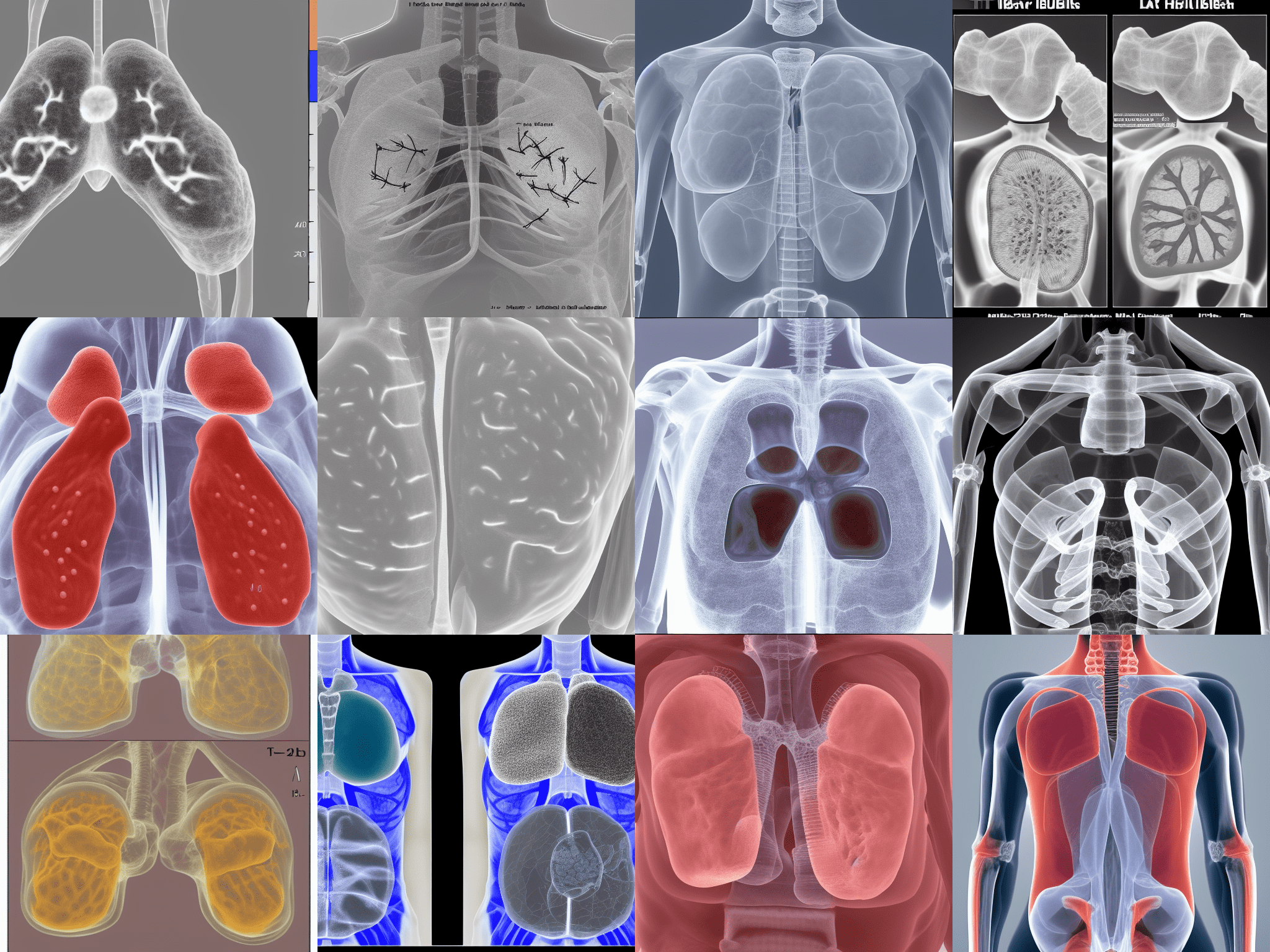}\label{fig:M0_abnormal}}
  \hfill
  \subfloat[M1-Abnormal (Adam8bit1)]{\includegraphics[width=0.24\textwidth]{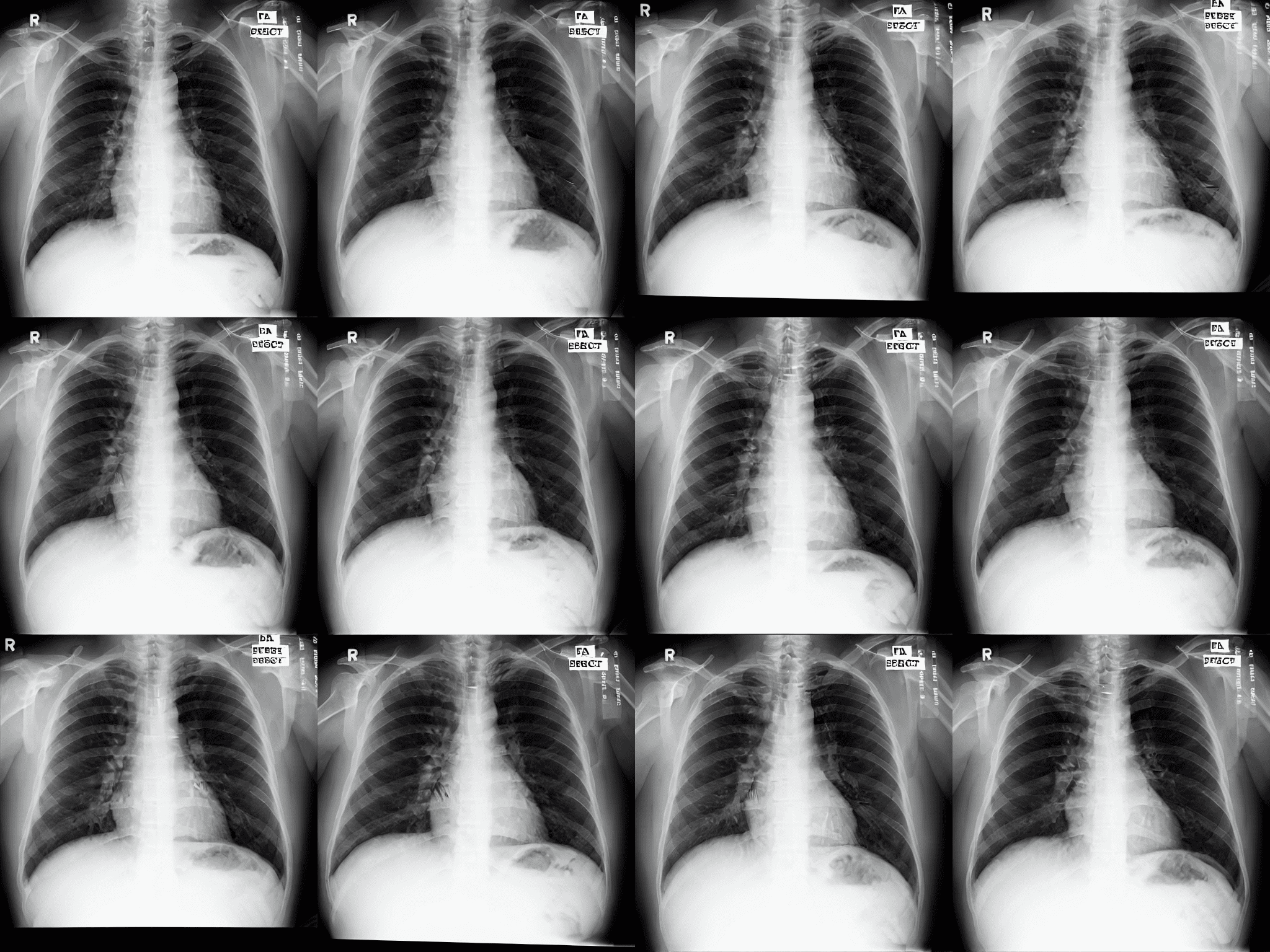}\label{fig:M1_abnormal}}
  \vfill
  \subfloat[M2-Abnormal (Adam8bit2)]{\includegraphics[width=0.24\textwidth]{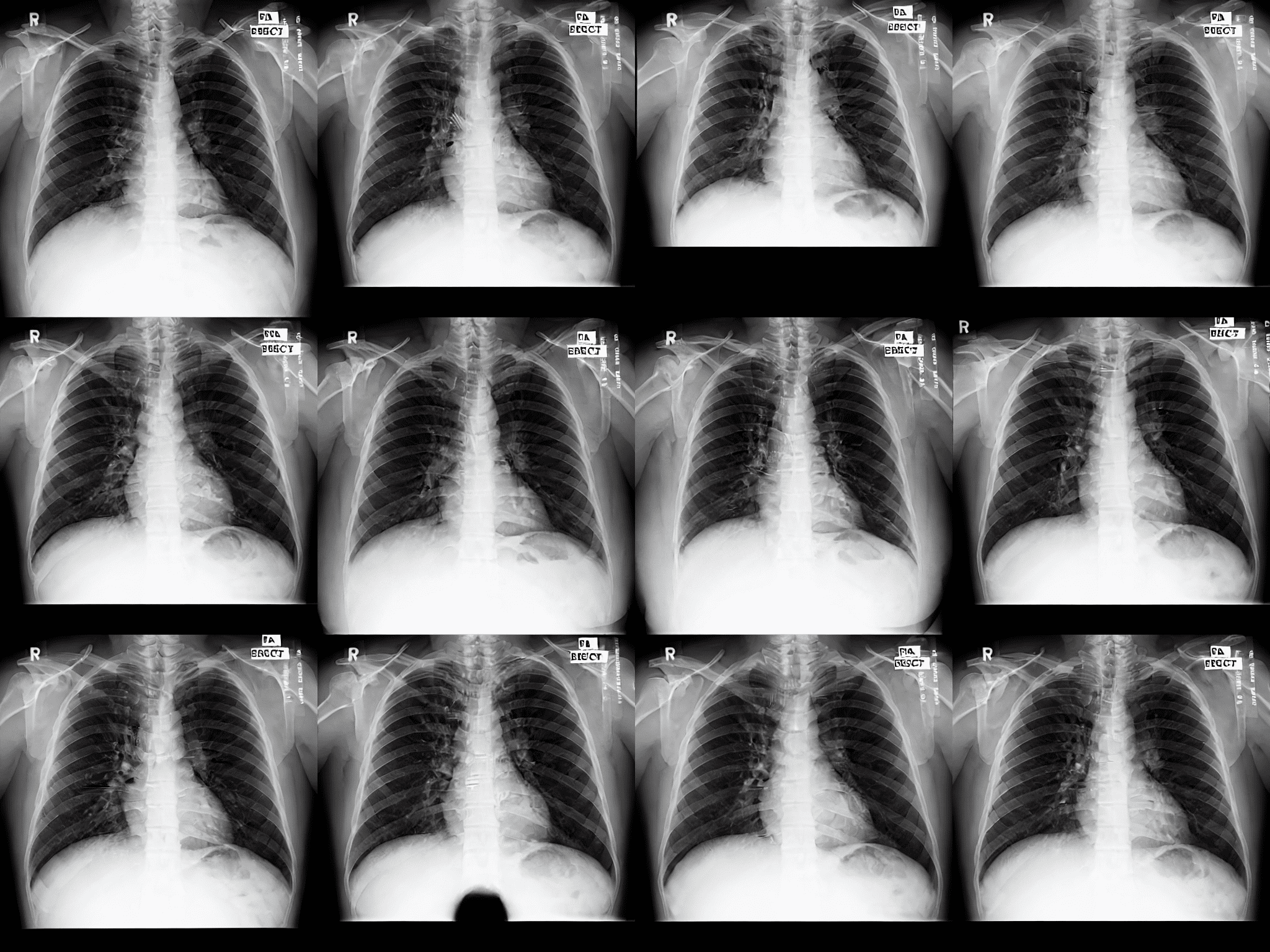}\label{fig:M2_abnormal}}
  \hfill
  \subfloat[M3-Abnormal (Adafactor)]{\includegraphics[width=0.24\textwidth]{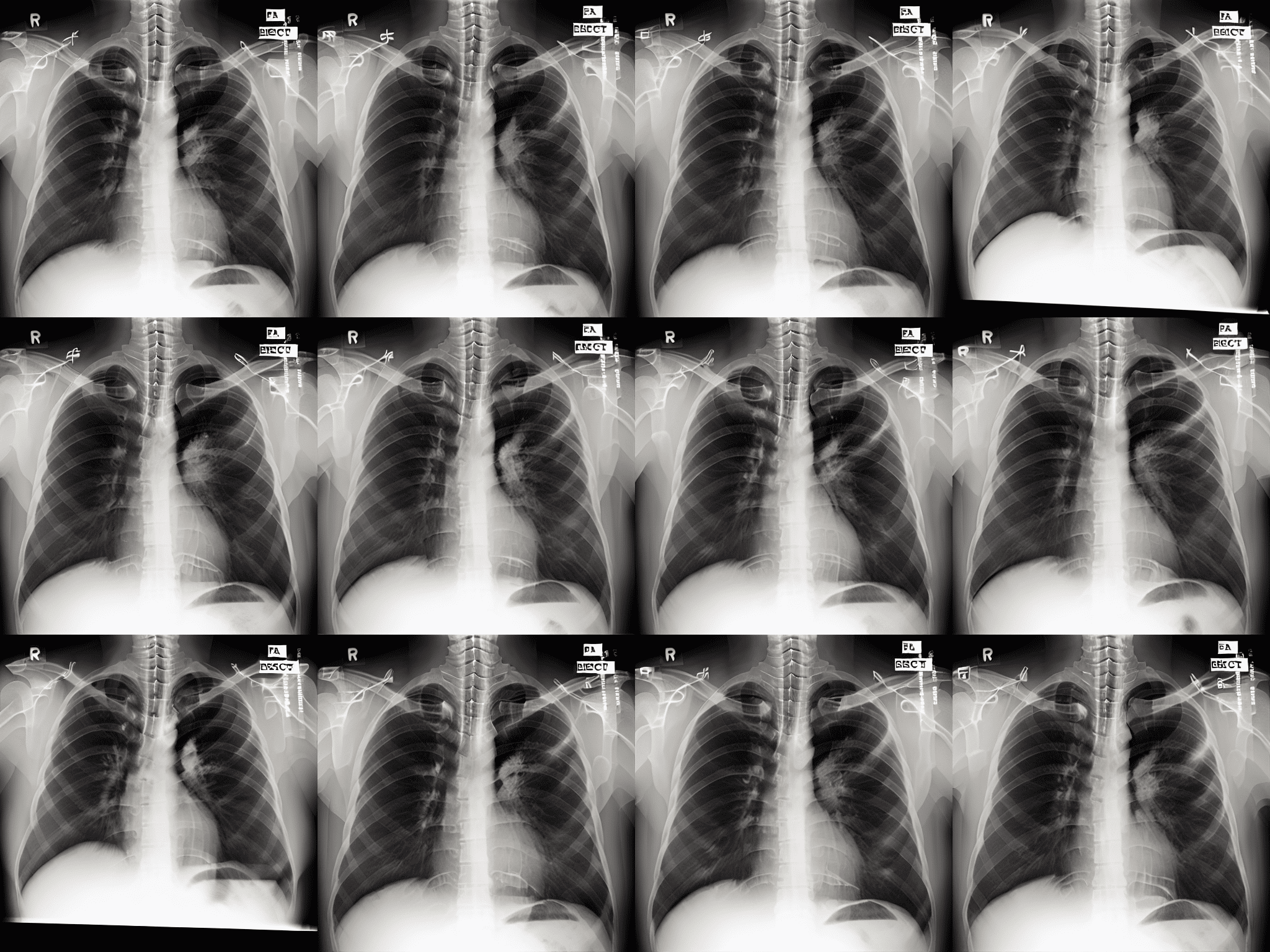}\label{fig:M3_abnormal}}
  \vfill
  \subfloat[M4-Abnormal (DAdaptSGD)]{\includegraphics[width=0.24\textwidth]{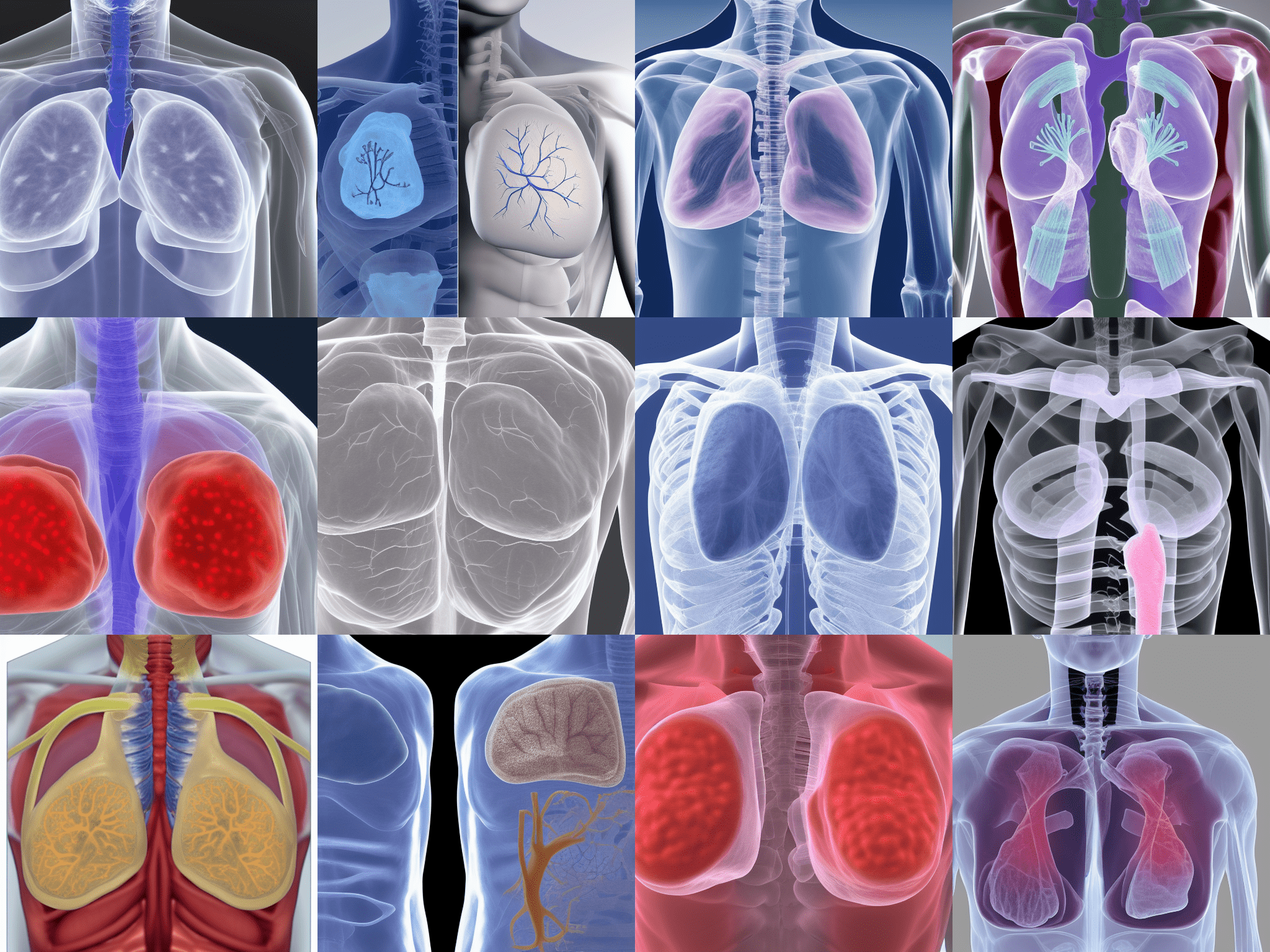}\label{fig:M4_abnormal}}
  \hfill
  \subfloat[M5-Abnormal (Prodigy)]{\includegraphics[width=0.24\textwidth]{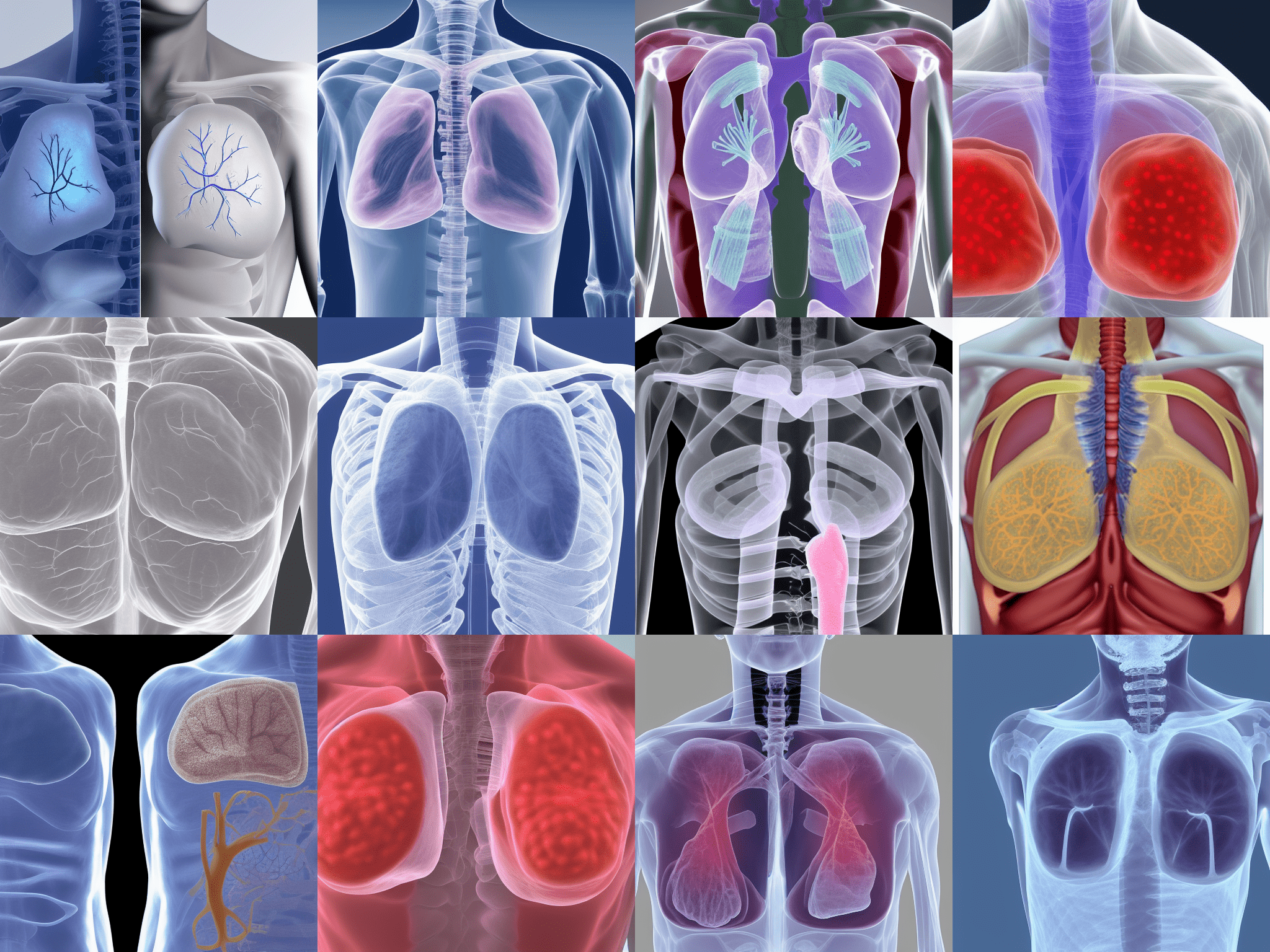}\label{fig:M5_abnormal}}
  \caption{Set of abnormal chest x-ray images generated by the models presented in Table \ref{tbl:configurations}. All models except M0 were fine-tuned for 100 epochs on a dataset of 30 chest x-ray images. All images were generated using the prompt \textit{"Human chest x-ray with tuberculosis. Bilateral miliary nodules with Right Middle Lobe infiltrate. Right pleural effusion"}.}
  \label{fig:abnormal_images}
\end{figure}

Figure~\ref{fig:normal_images} and Figure~\ref{fig:abnormal_images} present the sets of images generated for the normal and abnormal cases, respectively. All fine-tuned models (i.e., M1 through M5) were trained for 100 epochs. 
We presented the generated images to a medical doctor, who evaluated their level of realism using a five-level Likert scale from Very Unrealistic (1) to Very Realistic (5). Normal and abnormal images were evaluated separately for each model.

Table~\ref{tbl:results} presents the evaluation of the images made by the medical doctor regarding each model, with respect to their realism. Models M0, M3, M4, and M5 were evaluated as Very Unrealistic (1) for both image sets. Indeed, it can be empirically observed in Figures~\ref{fig:normal_images} and~\ref{fig:abnormal_images} that these models generate images that are quite different from a real x-ray image (Figure~\ref{fig:dataset_sample}). 
Additionally, models M4 and M5, in particular, presented a similar result to the foundational model M0 by generating more colorful and cartoon images.

Models M1 and M2 achieved better results, with M1 obtaining a Very Realistic (5) evaluation for normal cases and Average Realism (3) for abnormal cases. Both models utilize the Adam8bit optimizer, with M1 using the default configuration.
These results indicate that further experimentation could be conducted using the Adam8bit optimizer, as it demonstrated satisfactory performance. However, it is important to acknowledge the limitation of having only one evaluator.

\begin{table}[ht]
\centering
\caption{Results of the medical doctor's evaluation. Images generated by each model were evaluated using a five-level Likert scale, from Very Unrealistic (1) to Very Realistic (5).}
\begin{tabular}{@{}lccc@{}}
\toprule
Model           & \begin{tabular}[c]{@{}c@{}}Normal\\ (Figure~\ref{fig:normal_images})\end{tabular} & \begin{tabular}[c]{@{}c@{}}Abnormal\\ (Figure~\ref{fig:abnormal_images})\end{tabular} & Total \\ \midrule
M0 (Foundation) & 1                                                           & 1                                                             & 2     \\
M1 (Adam8bit1)  & 5                                                           & 3                                                             & 8     \\
M2 (Adam8bit2)  & 3                                                           & 2                                                             & 5     \\
M3 (Adafactor)  & 1                                                           & 1                                                             & 2     \\
M4 (DAdaptSGD)  & 1                                                           & 1                                                             & 2     \\
M5 (Prodigy)    & 1                                                           & 1                                                             & 2     \\ \bottomrule
\end{tabular}
\label{tbl:results}
\end{table}

\section{Conclusion}
\label{sec:final_considerations}

In this work, we presented an initial exploration of the impact of fine-tuning foundation models for generating medical images, focusing on chest X-rays. Our approach considers a Latent Diffusion Model as a base model and different optimizer configurations for the fine-tuning process. We generated a set of images using these models, considering both normal and abnormal cases.
Our experiments indicate that even the use of a small dataset for fine-tuning could generate images with satisfactory levels of realism. 
\red{However, our work has some limitations. The experiments were conducted with input from only one medical doctor and relied on visual inspection. Additional validation techniques could be applied, such as using qualitative metrics to assess the generated images and evaluating potential overfitting in the fine-tuned models.}

For future work, we consider further experimentation with different dataset sizes, evaluating the impact of different training times, and evaluating our model with more medical professionals. 
The Adam8bit optimizer, in particular, presented better results and could be used in such experiments. We also consider experimenting with additional prompts for abnormal cases, given the variety of conditions that could be presented in an x-ray.
Finally, we also consider the development of applications that will be beneficial in both healthcare and educational settings. One concept involves creating an application that allows teachers to use our method to generate personalized examples tailored to their students' needs. This approach can improve the teaching and learning experience by providing a more interactive and engaging environment.


\section*{Acknowledgments}

This study was partly funded by the Coordenação de Aperfeiçoamento de Pessoal de Nível Superior - Brazil (CAPES) - Finance Code 001, FAPERGS (RITE CIARS 22/2551-0000390-7), and by the Conselho Nacional de Desenvolvimento Científico e Tecnológico - Brasil (CNPq).
This paper was supported by the Ministry of Science, Technology, and Innovations, with resources from Law No. 8.248, dated October 23, 1991, within the scope of PPI-SOFTEX, coordinated by Softex, and published in the Residência em TIC 02 - Aditivo, Official Gazette 01245.012095/2020-56.
The authors would like to thank the medical doctor who participated in our experiments.




\bibliographystyle{IEEEtran}
\bibliography{bib}
%
%


\end{document}